\documentclass[12pt]{article}
\usepackage{graphicx}

\newcommand\pubnumber{Article 10  in eConf C1304143}
\newcommand\pubdate{\today}

\def\brera{INAF - Oss. Astron. di Brera.
Via E. Bianchi 46, I-23807 Merate, Italy}
\def\dado{Sydney Institute for Astronomy, The University of Sydney, NSW 2006, Australia}

\def\lara{APC Universit\'e Paris Diderot,  F-75205 Paris Cedex 13, France}
\def\ruben{INAF-IASF. Via Bassini 15,  I-20133 Milano, Italy}
\def\support{\footnote{Work supported by the PRIN-INAF 2011 "Four steps forward in 
understanding GRBs".}}

\def\Title#1{\begin{center} {\Large #1 } \end{center}}
\def\Author#1{\begin{center}{ \sc #1} \end{center}}
\def\Address#1{\begin{center}{ \it #1} \end{center}}

\newcommand\pubblock{\rightline{\begin{tabular}{l} \pubnumber\\
         \pubdate  \end{tabular}}}
\newenvironment{Abstract}{\begin{quotation}  }{\end{quotation}}
\newenvironment{Presented}{\begin{quotation} \begin{center} 
             PRESENTED AT\end{center}\bigskip 
      \begin{center}\begin{large}}{\end{large}\end{center} \end{quotation}}
\def\Acknowledgements{\bigskip  \bigskip \begin{center} \begin{large}
             \bf ACKNOWLEDGEMENTS \end{large}\end{center}}
             
\def\gsim{ \lower .75ex \hbox{$\sim$} \llap{\raise .27ex \hbox{$>$}} }
\def\lsim{ \lower .75ex\hbox{$\sim$} \llap{\raise .27ex \hbox{$<$}} }

\def\beq{\begin{equation}}
\def\eeq{\end{equation}}

\def\sw{{\it Swift}}
\def\fe{{\it Fermi}}
\def\ba{BATSE}
\def\cgro{{\it CGRO}}

\def\ep{$E_{\rm p}$}

\def\liso{$L_{\rm iso}$}
\def\eiso{$E_{\rm iso}$}
\def\ama{$E_{\rm p}-E_{\rm iso}$}
\def\yone{$E_{\rm p}-L_{\rm iso}$}
\def\ghi{$E_{\rm p}-E_{\gamma}$}

\def\th{$\theta_{\rm jet}$}

\def\thv{$\theta_{\rm view}$}
\def\tjet{$t_{\rm break}$}
\def\tpeak{$t_{\rm peak}$}
\def\egamma{$E_{\gamma}$}

\def\G{$\Gamma_{0}$}

\def\lisocom{$L'_{\rm iso}$}
\def\eisocom{$E'_{\rm iso}$}
\def\egcom{$E'_{\gamma}$}

\def\epcom{$E'_{\rm p}$}

\def\beq{\begin{equation}}
\def\eeq#1{\label{#1}\end{equation}}
\def\eeqn{\end{equation}}

\def\beqa{\begin{eqnarray}}
\def\eeqa#1{\label{#1}\end{eqnarray}}
\def\eeqan{\end{eqnarray}}

\let\bar=\overbar

\def\Dslash{\not{\hbox{\kern-4pt $D$}}}
\def\dslash{\not{\hbox{\kern-2pt $\del$}}}

\def\msb{{\bar{\ssstyle M \kern -1pt S}}}

\begin{document}
\begin{titlepage}
\pubblock

\vfill
\Title{On the relation between dynamics and geometry in GRBs}
\Author{ Giancarlo Ghirlanda\support\\
S. Campana, S. Covino, P. D'Avanzo\\ 
G. Ghisellini, A. Melandri, G. Tagliaferri}
\Address{\brera}
\Author{D. Burlon}
\Address{\dado}
\Author{L. Nava}
\Address{\lara}
\Author{R. Salvaterra}
\Address{\ruben}

\vfill
\begin{Abstract}
We estimate the initial bulk Lorentz factors \G\ for GRBs that show the onset of the afterglow in their optical light curves. We find that \G\ is strongly correlated with both the isotropic equivalent luminosity \liso\ and energy \eiso\ and, with a larger scatter, also with the rest frame peak energy \ep.  These new correlations allow us to interpret the spectral energy correlations $E_{\rm peak}-L_{\rm iso}$ ($-E_{\rm iso}$) as a sequence of $\Gamma_0$ factors. By accounting for the beaming effects, we find that the comoving frame properties of GRBs result clustered around typical values (e.g. $L^{\prime}_{\rm iso}\sim5\times 10^{48}$ erg/s). Moreover, it is theoretically predicted that there should be a link between the jet dynamics (\G) and its geometry (\th). Through a population synthesis code we reconstruct the \G\ and \th\ distributions and search for a possible link between them. We find that  \G\ and  \th\  in GRBs should have log--normal distributions and they should be anti correlated (i.e. $\theta_{\rm jet}^{2}\Gamma_0$=const). 
\end{Abstract}
\vfill
\begin{Presented}
Huntsville Gamma Ray Burst Symposium
GRB 2013\\
14-18 April 2013 Ð Nashville, Tennessee
\end{Presented}
\vfill
\end{titlepage}
\def\thefootnote{\fnsymbol{footnote}}
\setcounter{footnote}{0}

\section{Bulk Lorentz factor \G}

Gamma Ray Bursts (GRBs) are relativistic sources, as proposed theoretically (the compactness argument - e.g. \cite{lithwick}) and  proved by the observation of the transient radio scintillation in GRB 970508 \cite{frail}. The outflow bulk Lorentz factor $\Gamma$ increases initially with the distance from the source (acceleration phase) and then becomes constant (coasting phase - the prompt $\gamma$--ray emission is produced through internal shocks). Further out, due the interaction of the outflow with the circum-burst material, $\Gamma$  decreases (deceleration phase - the afterglow radiation is produced at the external shock). The deceleration time $t_{\rm dec}$ is typically defined when $\Gamma$ is halved with respect to its value during the coasting phase $\Gamma(t_{\rm dec})=\Gamma_0/2$ \cite{sari}. 
Observationally, this time is close to the peak of the afterglow light curve referred to as \tpeak. The estimate of $\Gamma_0$ is based on our present knowledge of the fireball dynamics. It depends on the density profile of the circum-burst medium $n\propto R^{-s}$ (where $R$ is the radial coordinate from the source) in such a way that $\Gamma_0\propto(E_{\rm k,iso}/n)^{{1}\over{(8-2s)}} t_{\rm peak}^{-{\left({3-s}\over{8-2s}\right)}}$ \cite{ghirlanda12,nava13} where $E_{\rm k,iso}$ is the kinetic energy left after the prompt emission (i.e. $E_{\rm k,iso}\sim E_{\rm iso}/\eta$, where $\eta$ is the prompt emission efficiency). $s=0$ ($s=2$) describes the typical ISM (wind) medium. 

\begin{figure}[htb]
\centering
\includegraphics[height=2.7in,trim=130 1 80 1]{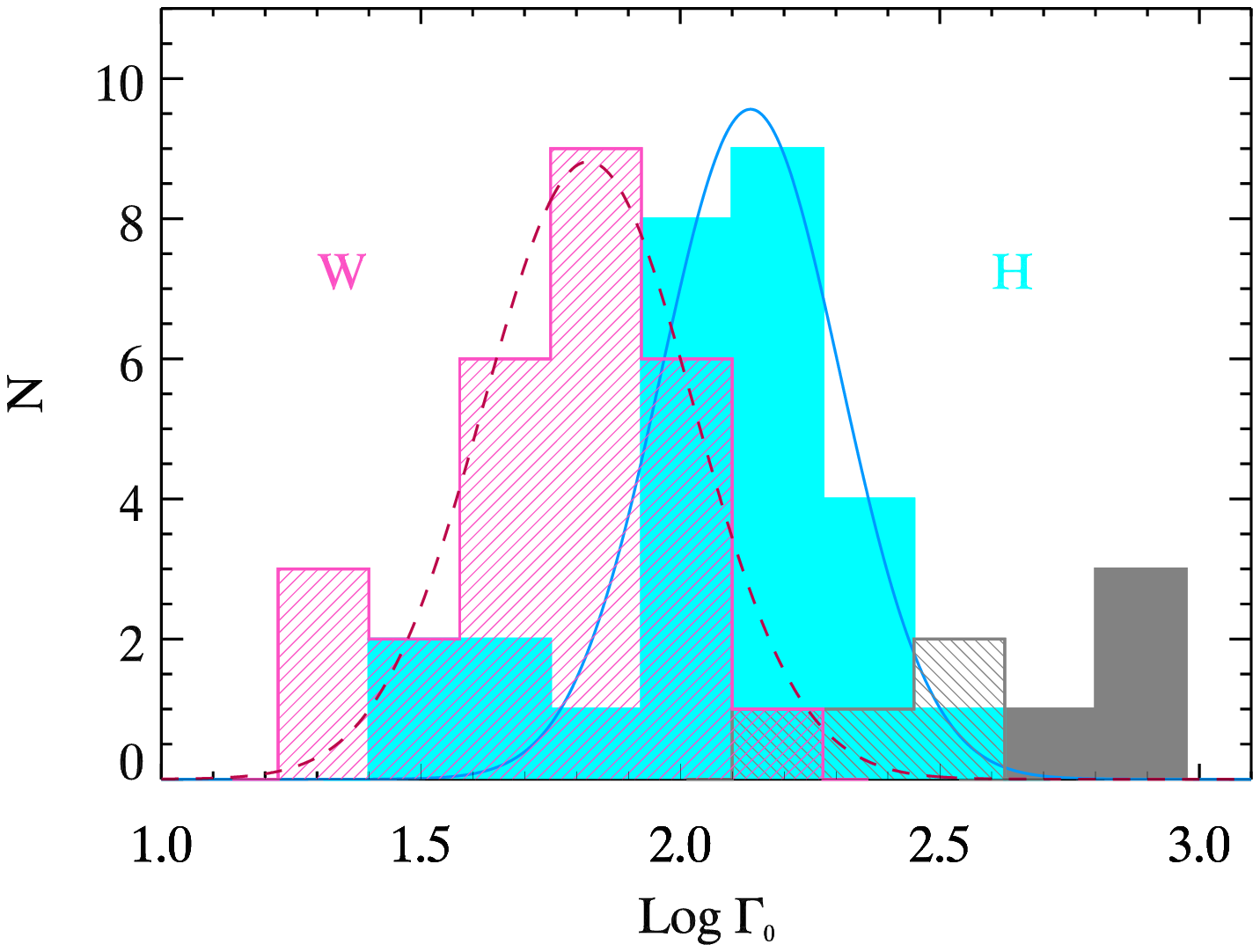}
\hskip 0.85in
\includegraphics[height=2.7in,trim=130 1 80 1]{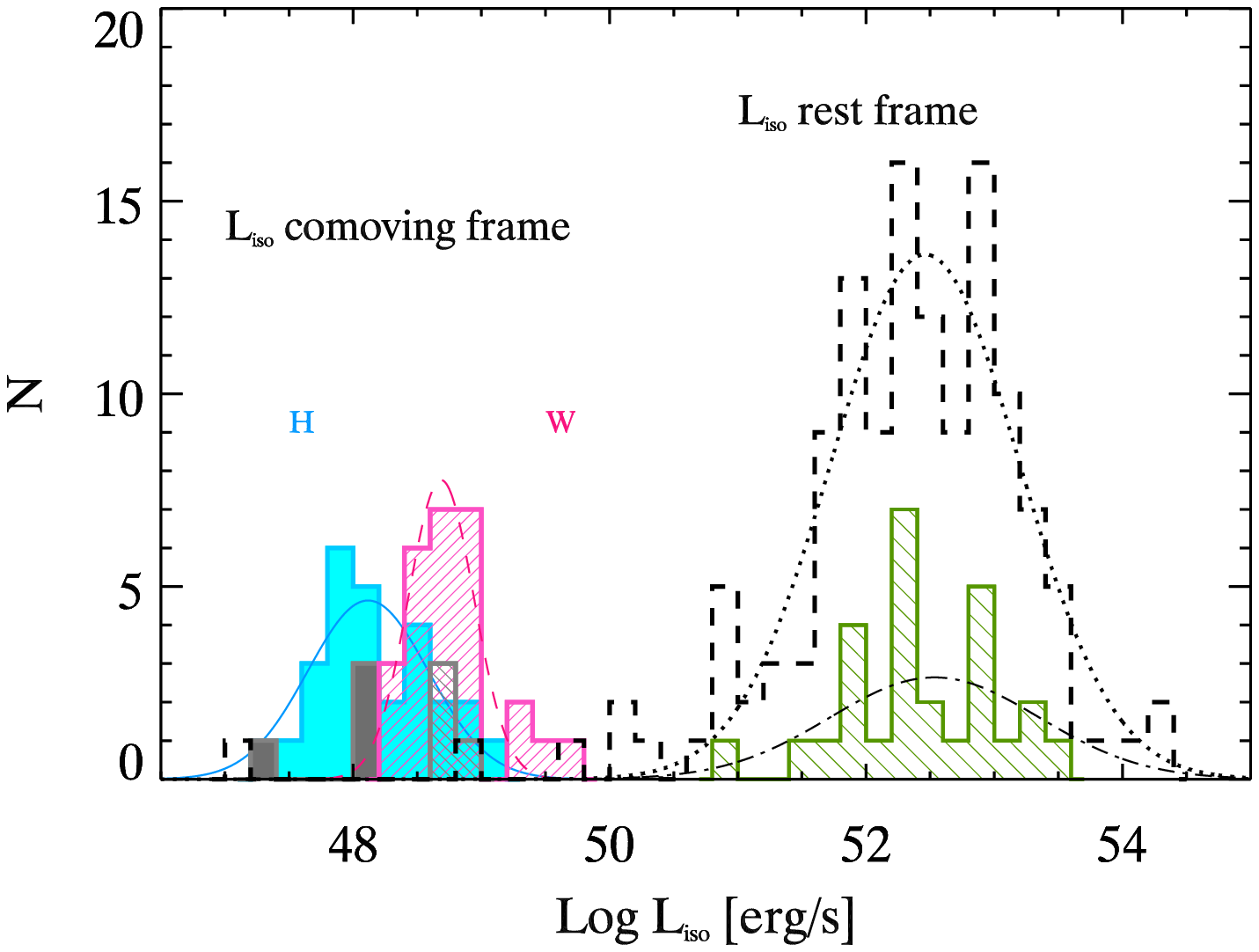}
\caption{\scriptsize Left: distributions of the bulk Lorentz factor \G\ in the ISM (H) and wind (W) case for the 27 GRBs of the sample of \cite{ghirlanda12} with a peak in the optical light curve. The solid and dashed lines are two fit with Gaussian functions. The 
three long (and one short) GRBs with a peak in the GeV light curve are shown with the grey (solid and hatched) histograms. Right: distributions of the isotropic equivalent luminosity \liso\ for the 131 GRBs with known redshift and well constrained \ep\ (necessary to compute \liso) and the GRBs of the sample of G12 with an estimate of \G\ (dashed and green hatched histogram respectively). On the left of the plot are shown the two distributions of the comoving frame luminosity \lisocom\ obtained in the H and W case. In the latter case the distribution of \lisocom\ is centered at 5$\times$10$^{48}$ with a dispersion of 0.07 dex (hatched pink histogram).}
\label{fg1}
\end{figure}

For the estimate of \G\  we need:  the deceleration peak time \tpeak, the redshift $z$ and the prompt emission isotropic energy \eiso. 
Assuming a typical value for the efficiency $\eta=0.2$ and density $n=3$ cm$^{-3}$ ($n=3.15\times10^{35}$ cm$^{-2}$) for the ISM (wind - corresponding to a wind velocity of 10$^3$ km s$^{-1}$ and 10$^{-5}$ M$_{\odot}$ yr$^{-1}$) profile\footnote{\G\ has a rather weak dependence on these parameters, i.e. \G$\propto(n\eta)^{-1/8}$ and \G$\propto(n\eta)^{-1/4}$ in the ISM and wind case.}, Ghirlanda et al. 2012 \cite{ghirlanda12} considered the 27 long GRBs with a peak in the optical light curve \footnote{X-ray light curve peaks were excluded because the emission in this energy range can be contaminated by an additional component than the afterglow - e.g. \cite{ghisellini09}.} plus the four GRBs (three long and one short event) with a peak in the GeV light curve (as detected by \fe/LAT). If the latter is interpreted as afterglow emission \G\ can be estimated as well from the \tpeak\ measured in the GeV light curve \cite{ghirlanda10,ghisellini,ackerman}. 

The {\it observed} distribution of \G\ is centered at 138 (66) in the ISM (wind) case (Fig.\ref{fg1} - left panel). The comoving frame distributions of the isotropic equivalent energy (\eisocom = \eiso/\G), of the peak energy (\epcom = \ep/(5\G/3)) and of the isotropic equivalent luminosity (\lisocom = \liso/(4\G$^{2}$/3)) cluster:  \eisocom\ and \epcom\ are centered at 1.4$\times$10$^{51}$ erg and $\sim$3 keV, respectively, for the ISM case and at 3$\times$10$^{51}$ erg and $\sim$6 keV for the wind case. These distributions are log--normal with a dispersion of about an order of magnitude. The comoving frame luminosity is much more clustered (especially in the wind case) around  a typical value of 5$\times$10$^{48}$ erg/s with a dispersion of only 0.07 dex as shown in Fig.\ref{fg1} (right panel). From these results, it seems that {\it GRBs have the same comoving frame properties (energetic/luminosity and peak energy).} 

The GRBs in the \cite{ghirlanda12} sample also show that \eiso$\propto$\G$^{2}$ (\liso$\propto$\G$^{2}$) and, with a wider dispersion, \ep$\propto$\G. 
Combining these newly found relations one retrieves the empirical correlations \ama\ (\yone) \cite{amati} (\cite{yonetoku}). Additionally, if one assumes that \th$^2$\G=const it is possible to derive the \ghi\ correlation (in the case of the wind circumburst medium \cite{nava06}) which involves the true GRB energy \egamma.

\section{The link between \G\ and  \th}

Another fundamental parameter of GRBs is the opening angle: both theoretical arguments and observational evidences have shown that GRBs have a jet with half opening angle \th. The present, still limited, sample of GRBs with an estimate of \th\ \cite{ghirlanda06} show that typically \th$\sim 3^{\circ}$. Consequently, \eiso\ (\liso) is only a proxy of the real GRB energetic (luminosity): \egamma=\eiso$(1-\cos\theta_{\rm jet}$). 
The estimate of \th\ is possible when an achromatic break in the afterglow light curve is observed between 1 and 10 days, typically. Based on the standard fireball model, this break should happen when the jet  bulk Lorentz factor is $\Gamma\sim1/\theta_{\rm jet}$ so that \th$\propto(n/E_{\rm k,iso})^{\frac{1}{(8-2s)}}$\tjet$^{\left(\frac{3-s}{8-2s}\right)}$  \cite{chevalier}.  Notably, one can derive the product \th\G$\propto\left(t_{\rm jet}/t_{\rm peak}\right)^{\frac{3-s}{8-2s}}$ which is independent of the redshift and other parameters. 

The estimate of \G\ requires  early time observations of the optical (or GeV) emission to measure the deceleration peak time. The right hand side of the \G\ distribution (left panel in Fig.\ref{fg1}) is limited by the difficulty of following the afterglow emission at early times thus preventing the estimate of large \G\ values.  Indeed, the larger values of \G\ have been derived from the \fe/LAT light curve thanks to the monitoring of the afterglow emission from the very beginning of the burst. 

The estimate of \th\ requires to follow the afterglow emission up to few days after the burst explosion. The measurement of early \tjet\ is complicated by the presence, more often in the X--ray light curve, of multiple breaks or by the contamination of late prompt emission, e.g. \cite{ghisellini09}. This limits our knowledge of the \th\ distribution towards the low \th\  values. On the other side, the distribution of \th\ towards large angles (corresponding to measurements of late \tjet) is hampered by the possible contamination, at late times,  by the SN emission and the host galaxy . 

Therefore, the presently known distributions of \th\ and \G\  could be subject to observational biases. Moreover, as explained above, a unifying explanation of the  \ama, \yone\ and \ghi\ correlations is possible if the newly found correlations between \G\ and \eiso,\liso\ and \ep\ are used together with the assumption that there is a link between the GRB jet dynamics (\G) and its geometry (\th). 

Motivated by these considerations,  we built a population synthesis code \cite{ghirlanda13} aimed at deriving the intrinsic distributions of \th\ and \G\ and at exploring the possible correlation between them. ``Intrinsic'' properties means those of the entire population of GRBs: due to the collimated nature of these sources, all we know observationally about them is related to the population of sources that we detect from the Earth, i.e. these are the bursts pointing to us. For each GRB with a given \th\ pointing to the Earth there are $(1-\cos\theta_{\rm jet})^{-1}$ objects pointing elsewhere.  We account for this effect in our code considering also the viewing angle \thv\ of the observer with respect to the jet axis. 

 We simulate a population of GRBs distributed in the redshift range $z\in[0,10]$ with a probability density (evolving in redshift as found by \cite{salvaterra}) given by the GRB formation rate of \cite{li}. We assign to each simulated burst a \G\ and \th\ extracted from  parametric distributions which we aim to constrain. We further assume that each burst is observed with a viewing angle \thv\ (following a probability density $\propto\sin$\thv). We assume that all bursts have a prompt emission spectrum described by a Band function. 

\begin{figure}[htb]
\centering
\includegraphics[height=2.5in,trim=130 1 80 1]{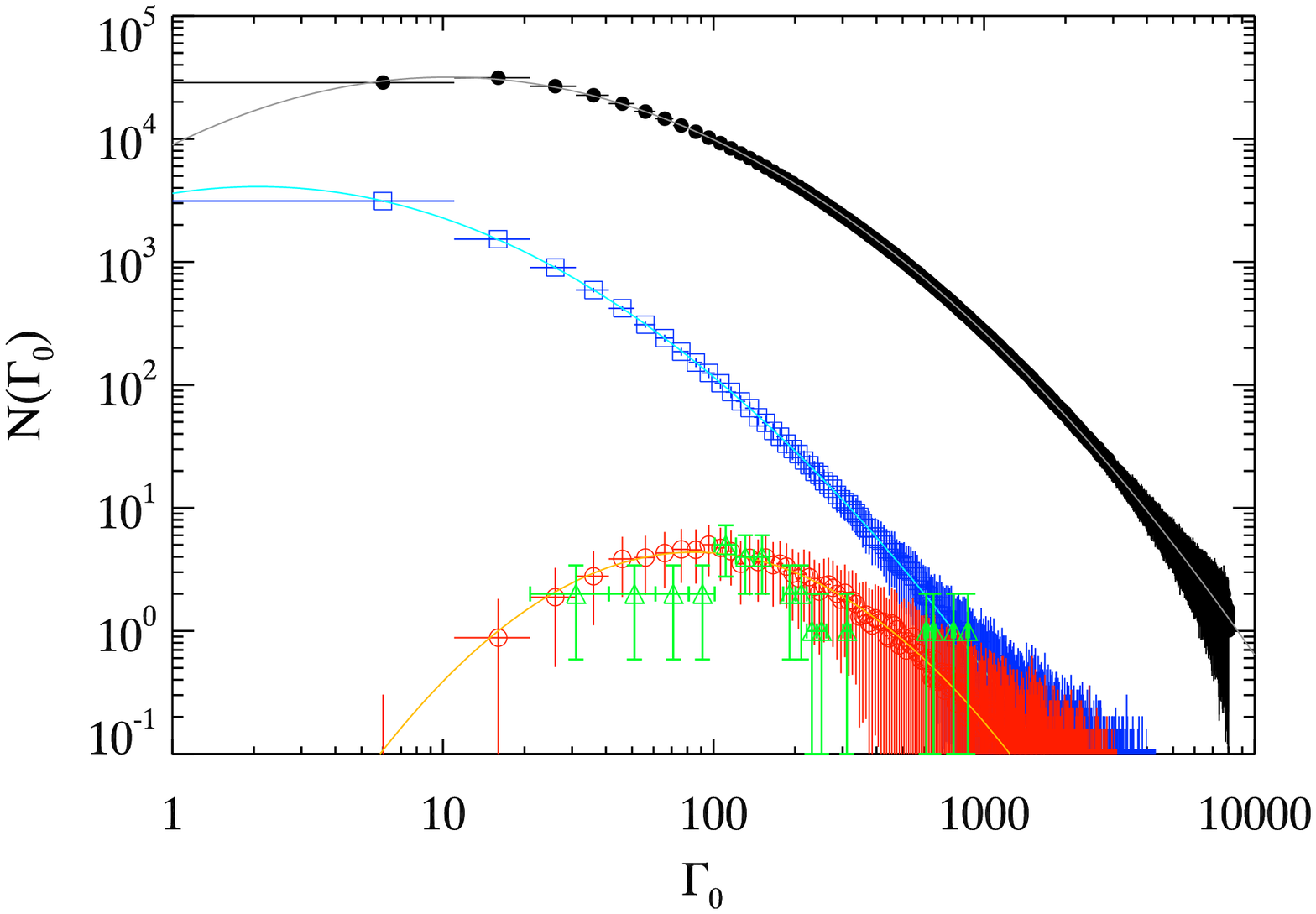}
\hskip 0.85in
\includegraphics[height=2.5in,trim=130 1 80 1]{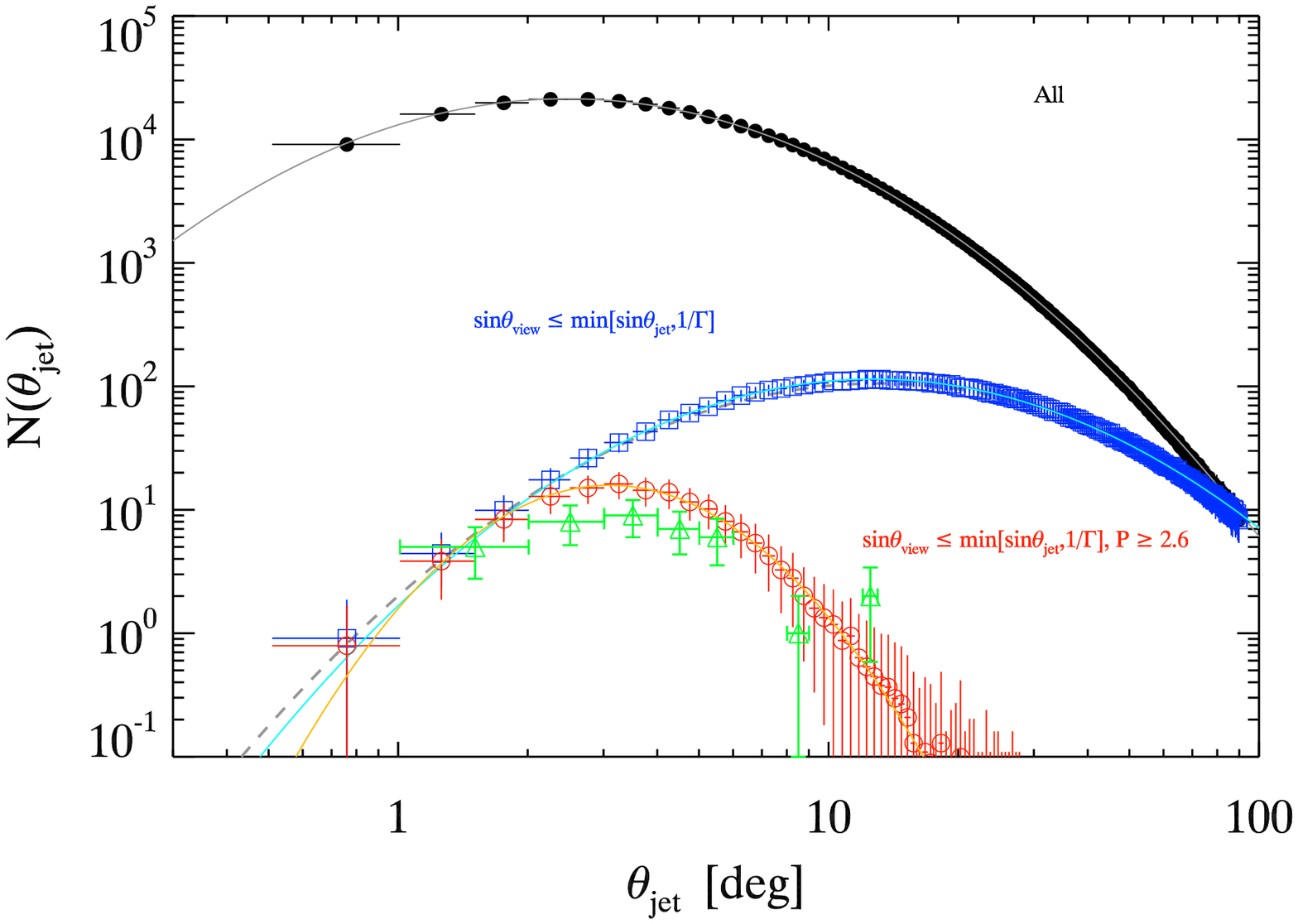}
\caption{\scriptsize Left: distribution of the bulk Lorentz factor \G\ of the simulated bursts. The (black) filled circles distribution represents the entire population, i.e. burst pointing in all directions distributed up to $z=10$. The (blue) open squares show the distribution of bursts pointing towards the earth (\thv$\le$min[\th,arsin(1/\G)]). Among the latter the (red) open circles are the simulated bursts equivalent to the \sw\ bright sample. Just for the comparison the distribution of known \G\ is shown with the (green) open triangles. Right: distribution of the jet half opening angles (same symbols as described above for the left panel). All lines show the fits with log--normal distributions (parameters of the fits can be found in \cite{ghirlanda13}.}
\label{fg2}
\end{figure}

Based on the results of G12 we assume that all GRBs have the same comoving frame \egcom\ and \epcom. For each simulated burst we compute the rest frame true energy \egamma=\egcom\G\ and the peak energy \ep=\epcom\G. The isotropic equivalent energy is \eiso=\egamma/($1-\cos$\th) if 1/\G$\le\sin$\th, i.e. if the collimation is prevailing over the relativistic beaming, or \eiso=\egamma(1+$\beta_{0}$)\G$^2$  (where 
$\beta_{0}=v_{0}/c$) in the opposite case, i.e. 1/\G$>\sin$\th. The prompt emission of GRBs observed at an angle \thv$>$\th\ can still be visible if their beaming angle is 1/\G$>\sin$\th. Finally, we evaluate the flux in a given energy range $\Delta$E in order to compare the simulated bursts with the observed distributions of GRBs observed by \cgro/Batse and \fe/GBM. 

Among the simulated bursts (black filled circles in Fig.\ref{fg2}) we extract the sub--population of GRBs pointing to the Earth (blue open squares in Fig. \ref{fg2}) and among these the bright sample, i.e. those with a peak flux $\ge$2.6 ph cm$^{-2}$ s$^{-1}$. The latter selection corresponds to the bright flux cut used to define the \sw\ complete sample (BAT6 - \cite{salvaterra}. We want to reproduce with the simulated bursts the \ama\ correlation (in slope, normalization and scatter) as defined with the complete \sw\ sample BAT6 \cite{nava12} and the flux distributions of  \ba\ and \fe\ GRBs. 

Through this code we can derive the intrinsic distributions of \G\ and \th. We find (Fig.\ref{fg2}) that \G\ and \th\  should have log--normal distributions (simple power law distributions fail to reproduce some of the observational constraints -- see \cite{ghirlanda13}). The black points in Fig.\ref{fg2} show the entire population of GRBs which should have a typical \th\ and \G\ of 4.5$^{\circ}$ and 90, respectively (right and left panel of Fig.\ref{fg2}). However, from the Earth we can only sample the distributions of the objects pointing to the Earth (blue open circles in Fig.\ref{fg2}) thus missing very small \th\ (right panel) and large \G\ (left panel). Despite this natural bias, we note that by considering only the bright sample of GRBs (red open circles in Fig.\ref{fg2}) we are sampling, on average, reasonably well the intrinsic distributions of \th\ and \G (black symbols): indeed the red and black distributions have similar peak values. As an a--posteriori check of our code, we show the distributions of the real bursts with measured \G\ and \th\ (green open triangles in both panels of Fig.\ref{fg2}).  

In other words, these results also suggest that if we were able to measure \th\ and \G\ for fainter  bursts we would sample the population of the objects pointing to the Earth which is far from being representative of the intrinsic distributions, i.e. we would be more subject to the natural selection effect of the viewing angle. On the other hand, our results suggest that, through the bright GRB population explored so far, we have accessed only partly the spectral--energy correlation plane. If, the scatter of the \ama\ correlation is due to \th, it should be larger than the present value (0.45 dex - \cite{nava12}) and it should be highly asymmetric (i.e. with more data points on the left--hand side of the \ama\ correlation) due to the log--normal distribution of \th. 

The jet break, observed so far in a still limited sample of bursts, is expected when the bulk Lorentz factor, which is decreasing during the afterglow phase, becomes comparable to 1/\th. However, if the jet starts with a value \G$>$1/\th\ it will never satisfy this condition. These are the GRBs that will never show a jet break in their afterglow light curve. This effect could  explain the few GRBs  (we estimate $\sim$6\%) that, despite being observed up to extremely late times, do not present any jet break in their optical light curve (see \cite{ghirlanda13} for more details). 

Within our modeling we could also derive the real rate of GRBs (i.e. number of objects Gpc$^{-3}$ yr$^{-1}$ pointing in all directions) as a function of the redshift. This number can  be compared with that of SNIbc: we find, similarly to other estimates e.g. \cite{guetta},  that the fraction of SNIbc producing a GRB should be $\sim$0.3\%. Our modeling, differently from past estimates that adopted an average value of the beaming correction to estimate the true GRB rate, is based on the intrinsic population of GRBs reproduced by our population synthesis code (right panel of Fig.\ref{fg2}). 

Finally, according to our code, there should be a relation between \th\ and \G. If we simply assume that these two parameters are uncorrelated, we cannot reproduce all the observational constraints we have. In particular we would find a distribution of bursts in the \ama\ plane which define a steeper (than  observed )\ama\ correlation. We tested and confirmed the hypothesis that the larger is (on average) the \G\ the smaller is \th, so that \th$^2$\G=const. This result, obtained for the first time through a population synthesis code, confirms what is expected from magnetically acceleration models of GRBs jets \cite{tchekhovskoy,komissarov}.

\Acknowledgements
This research has been supported by a PRIN-INAF 2011 grant.

\end{document}